\documentclass[11pt]{article}
\usepackage{amsfonts,latexsym}

\def\underset#1#2{\mathrel{\mathop{#2}\limits_{#1}}}
\def\overset{\stackrel}
 
\makeatletter
\@addtoreset{equation}{section}
\makeatother
\renewcommand{\theequation}{\thesection.\arabic{equation}}

\newfont{\blackb}{msbm10 scaled\magstep1}
 
\newcounter{subequation}[equation]
\makeatletter

\expandafter\let\expandafter
\reset@font\csname reset@font\endcsname

\def\subeqnarray{\arraycolsep1pt
    \def\@eqnnum\stepcounter##1{\stepcounter{subequation}%
        {\reset@font\rm(\theequation\alph{subequation})}}
\jot5mm     \eqnarray}

\def\subarray{\arraycolsep1pt
    \def\@eqnnum\stepcounter##1{\stepcounter{subequation}%
        {\reset@font\rm(\alph{subequation})}}
\jot5mm     \eqnarray}

\makeatother

\newfont{\calig}{cmsy10 scaled\magstep1}

\def\text#1{\hbox{#1}}
 
\newtheorem{theorem}{Theorem}[section]

\newtheorem{remark}{Remark}[section]
\newtheorem{corollary}{Corollary}[section]
\newtheorem{lemma}{}
\newtheorem{definition}{Definition}[section]
 
\newcommand{\qed}{\hfill$\Box$}

\def\non{\nonumber\\}
\def\be{\begin{equation}}
\def\ee{\end{equation}}
\def\ben{\begin{displaymath}}
\def\een{\end{displaymath}}
\def\baa{\begin{eqnarray}}                                                    
\def\eaa{\end{eqnarray}} 
\def\baan{\begin{eqnarray*}}
\def\eaan{\end{eqnarray*}}  
\def\ba{\begin{array}}
\def\ea{\end{array}} 
\def\a{\alpha}
\def\b{\beta}
\def\g{\gamma}

\def\ka{\kappa}
\def\l{\lambda}
\def\L{\Lambda}

\def\ph{\phi}
\def\r{\rho}
\def\s{\sigma}

\def\Th{\Theta}
\def\o{\omega}
\def\Om{\Omega}

\def\half{\textstyle{\frac12}}
\def\ie{{\it i.e.}}
\def\eg{{\it e.g.}}
\def\la{\label}
\def\Ref{\ref}
\def\c{\cite}
\def\f{\frac}
\def\p{\partial}

\def\Ref#1{(\ref{#1})}

\def\0{S}
\def\1{T}

\def\tPsi{{\widetilde{\Psi}}}
\def\tpsi{{\widetilde{\psi}}}

\def\str{{\rm str}}
\def\endo{{\rm End}}

\def\hb{{\widehat{\b}}}
\def\vm{{\vec{\mu}}}
\def\vz{{\vec{z}}}
\def\vph{\varphi}
\def\Ka{{\mathcal{K}}}
\def\wtt {{\widetilde{t}}}
\begin{document}

\begin{titlepage}

\begin{center}

\phantom.
\vskip2.5cm

{\huge \bf Bethe vectors of the \(osp(1|2)\) 
Gaudin model}\\[5mm]

\bigskip

{\sc P. ~P.~Kulish
\footnote{E-mail address: kulish@pdmi.ras.ru \,;
On leave of absence from Steklov Mathematical Institute, 
Fontanka 27, 191011, St.Petersburg, Russia.} 
and N. ~Manojlovi\'c 
\footnote{E-mail address: nmanoj@ualg.pt}\\ 
\medskip
{\it \'Area Departmental de Matem\'atica, F. C. T., 
Universidade do Algarve\\ Campus de Gambelas, 8000 Faro, 
Portugal}}\\ 

\vskip2.5cm 

\end{center}

\begin{abstract}
The eigenvectors of the $osp(1|2)$ invariant Gaudin hamiltonians 
are found\break\hfil 
using explicitly constructed creation operators. Commutation
relations between the creation operators and the generators of the 
loop superalgebra are calculated. The coordinate representation of
the Bethe states is presented. The relation between the 
Bethe vectors and solutions to the Knizhnik-Zamolodchikov equation 
yields the norm of the eigenvectors.
\end{abstract} 

\end{titlepage}

\clearpage \newpage


\setcounter{page}1

\section{Introduction} 

Classifying integrable systems solvable in the framework of the 
quantum inverse scattering method \c{F95,KS} by underlying dynamical 
symmetry algebras, one could say that the Gaudin models are the 
simplest ones being related to loop algebras and classical 
$r$-matrices. More sophisticated solvable models correspond to 
Yangians, quantum affine algebras, elliptic quantum groups, etc.  

Gaudin models \c{GB} are related to classical $r$-matrices, 
and the density of Gaudin hamiltonians 
\be \la{Gh}  
H^{(a)} = \sum_{b \neq a} r_{ab}(z_a - z_b) 
\ee 
coincides with the $r$-matrix. 
Condition of their commutativity $[ H^{(a)} , H^{(b)} ]$ $= 0$
is nothing else but the classical Yang-Baxter equation (YBE). 

The Gaudin models (GM) associated to classical  $r$-matrices of 
simple Lie algebras were studied in many papers 
(see [3-9] and references therein).
The spectrum and eigenfunctions were found using different methods
(coordinate and algebraic Bethe Ansatz \c{GB,S87}, separated 
variables \c{S99}, etc.). A relation to the Knizhnik-Zamolodchikov 
(KZ) equation of conformal filed theory was established [7-9].
 
There are additional peculiarities of Gaudin models related to
classical $r$-matrices based on Lie superalgebras due to 
$Z_2$-grading of representation spaces and operators. The study
of the $osp(1|2)$ invariant Gaudin model corresponding to the 
simplest non-trivial super-case of the $osp(1|2)$ invariant 
$r$-matrix \c{K85} started in \c{BM}. The spectrum of the $osp(1|2)$ 
invariant Gaudin hamiltonians $H^{(a)}$  was given, antisymmetry 
property of their eigenstates was claimed, and a two site model 
was connected with some physically interesting one 
(a Dicke model).  

The creation operators used in the $sl(2)$ GM coincide 
with one of the $L$-matrix entry \c{GB,S87}. However, in the 
\(osp(1|2)\) case, as we will show, the creation operators 
are complicated polynomials of two generators $X^+(\l)$ and 
$v^+(\mu)$ of the loop superalgebra. We introduce $B$-operators 
by a recurrence relation (Section 3). Acting on the lowest spin 
vector the $B$-operators generate exact eigenstates of the 
Gaudin hamiltonians $H^{(a)}$, provided Bethe equations are 
imposed on parameters of the states.
Furthermore, the recurrence relation is solved 
explicitly and the commutation relations between the 
$B$-operators and the generators of the loop superalgebra 
${\cal L}(osp(1|2))$ are calculated. We prove that the constructed 
states are lowest spin vectors of the global finite dimensional 
superalgebra $osp(1|2)$, as it is the case for many invariant 
quantum integrable models \c{TF}. Moreover, a striking coincidence
between the spectrum of the $osp(1|2)$ invariant Gaudin 
hamiltonians of spin $s$ and the spectrum of the hamiltonians 
of the $sl(2)$ GM of the integer spin $2s$ is found (Section 3).

A connection between the $B$-states, when the Bethe equations are 
not imposed on their parameters, of the Gaudin models for simple 
Lie algebras to the solutions to the Knizhnik-Zamolodchikov 
equation was established in the papers \c{BF, FFR}. 
An explanation of this connection based on Wakimoto modules 
at critical level of the underlying affine algebra was given in 
\c{FFR}. An explicit form of the Bethe vectors in the coordinate 
representation was given in both papers \c{BF, FFR}. 
The coordinate Bethe Ansatz for the $B$-states of 
the $osp(1|2)$ GM is obtained in our paper as well. Using 
commutation relations between the $B$-operators 
and the transfer matrix $t(\l)$ we demonstrate algebraically 
that explicitly constructed $B$-states yield a solution to the 
KZ equation (Section 4). This connection permits us to calculate 
the norm of the eigenstates of the Gaudin hamiltonians.
An analogous connection is expected between quantum $osp(1|2)$ 
spin system related to the graded YBE \c{K85,M} and quantum 
KZ equation following the lines of \c{TV}. 

The norm and correlation functions of the $sl(2)$ invariant 
GM were evaluated in \c{S99} using Gauss factorization 
of a group element and Riemann-Hilbert problem. The study of this 
problem for the GM based on the $osp(1|2)$ Lie superalgebra is 
in progress. Nevertheless we propose a formula for the scalar 
products of the Bethe states which is analogous to the $sl(2)$ 
case (Conclusion).

  
\section{$OSp(1|2)$-invariant $R$-matrix}   

Many properties of the Gaudin models can be obtained as a 
quasi-classical limit of the corresponding quantum spin systems 
related to solutions $R(\l; \eta)$ to the YBE. In the 
quasi-classical limit $\eta \to 0$
\ben 
R(\l; \eta) = I + \eta r(\l) + {\cal O}(\eta^2), 
\een
some relations simplify and therefore can be solved explicitly 
providing more detailed results for the GM. 

The graded Yang-Baxter equation \c{KS, K85} differs from the usual 
YBE by some sign factors due to the embedding of $R$-matrix into the 
space of matrices acting on the $Z_2$-graded tensor product 
$V_1 \otimes V_2 \otimes V_3$. At this point our aim is to reach 
fundamental $osp(1|2)$ invariant solution. The rank of the 
orthosymplectic Lie algebra $osp(1|2)$ is one and its dimension is 
five. The three even generators are $h, X^+, X^-$ and 
the two odd generators are $v^+, v^-$ \c{Ritt}. The (graded) 
commutation relations between the generators are 
\be \la{osp}
\ba{cc}
\left[ h, X^{\pm} \right] = \pm 2 X^{\pm} \;,&
\left[ X^+, X^- \right] = \, h \;, \\ 
\left[ h, v^{\pm} \right] = \pm v^{\pm} \;, &  
\left[ v^+, v^- \right] _+ = - h \;, \\
\left[ X^{\mp}, v^{\pm} \right] = \, v^{\mp} \;, &  
\left[ v^{\pm}, v^{\pm} \right] _+ = {\pm} 2 X^{\pm} \;, 
\ea
\ee
together with $\left[ X^{\pm}, v^{\pm} \right] = 0$. Notice that the 
generators $h$ and $v^{\pm}$ considered here \Ref{osp} differ by 
a factor of 2 from the ones used in \c{Ritt,K85}. Thus the Casimir 
element is 
\baa \la{Cas}  
c_2 &=& h^2 + 2 \left(X^+ X^- + X^- X^+ \right) 
+ \left(v^+ v^- - v^- v^+ \right)  \nonumber\\
&=& h^2 - h + 4 X^+ X^- + 2 v^+ v^-  \,.
\eaa
For further comparison with the $sl(2)$ Gaudin model \c{GB,S87} 
and due to the chosen set of generators \Ref{osp} we parameterize 
the finite dimensional irreducible representations $V^{(l)}$ of the 
$osp(1|2)$ Lie superalgebra by an integer $l$, so that their 
dimensions $2l + 1$ and the values of the Casimir element 
\Ref{Cas} $c_2 = l(l+1)$ coincide with the same characteristics 
of the integer spin $l$ irreducible representations of $sl(2)$. 

The fundamental irreducible representation $V$ of $osp(1|2)$ 
is three dimensional. We choose a gradation 
of the basis vectors $e_j;\, j=1,2,3$ to be $(0, 1, 0)$. 

The invariant $R$-matrix is a linear combination \c{K85} 
\be \la{R} 
R= \l \left( \l+\frac {3\eta}2 \right )I + \eta
\left( \l+\frac {3\eta}2 \right) {\cal P} - \eta \, \l \, K \,, 
\ee 
of the three $OSp(1|2)$ group invariant operators 
$[ g\otimes g, X] = 0$, $g \in OSp(1|2)$, 
$X \in {\endo} \, (V \otimes V)$,  
acting on $V \otimes V$: the identity $I$, the graded 
permutation $\cal P$ and a rank one projector $K$. 
In the equation \Ref{R} $\l$ is the spectral parameter, 
and $\eta$ is a quasi-classical parameter. 
 
The $L$-operator of the quantum spin system on a 
one-dimensional lattice with $N$ sites coincides with 
$R$-matrix acting on a tensor product 
$V_0 \otimes V_a$ of auxiliary space $V_0$  and 
the space of states at site $a = 1, 2,\dots N$ 
\be \la{Lq} 
L_{0a}(\l - z_a) = R_{0a} (\l - z_a) \,, 
\ee 
where \(z_a\) is a parameter of inhomogeneity 
(site dependence) \c{F95,KS}. 
Corresponding monodromy matrix $T$ is an ordered product 
of the $L$-operators 
\be \la{Tq} 
T(\l ; \{z_a\}_1^N) = L_{0N}(\l - z_N) \dots  
L_{01}(\l - z_1) = 
\underset{\longleftarrow}{\prod_{a=1}^{N}}
L_{0a} (\l - z_a) \,.  
\ee
The commutation relations of the $T$-matrix entries follow 
from the FRT-relation \c{F95} 
\be \la{RTT}  
R_{12}(\l - \mu)T_1(\l)T_2(\mu) = 
T_2(\mu)T_1(\l)R_{12}(\l - \mu) \,.  
\ee 
Multiplying \Ref{RTT} by $R_{12}^{-1}$ and taking the 
super-trace over $V_1 \otimes V_2$, one gets 
commutativity of the transfer matrix 
\be \la{tq} 
t(\l) = \sum_j (-1)^{j+1}T_{jj}(\l ; \{z_a\}_1^N)
= T_{11} - T_{22} + T_{33}  
\ee 
for different values of the spectral parameter 
$t(\l)t(\mu) = t(\mu)t(\l)$.

The choice of the $L$-operators \Ref{Lq} corresponds to 
the following space of states of the $osp(1|2)$-spin system  
\ben
{\cal H} = \underset {a=1}{\overset{N}{\otimes}} V_a \;. 
\een
The eigenvalue of the transfer matrix $t(\l)$ 
in this space is \c{K85} 
\baa 
\la{lq} 
\L (\l ; \{\mu_j\}_1^M) &=& 
\a_1^{(N)}(\l;\{z_a\}_1^N) \prod_{j=1}^M S_1(\l - \mu_j) 
-\a_2^{(N)}(\l; \{z_a\}_1^N) \times
\non
&\times& \prod_{j=1}^M S_1 \left( \l - \mu_j + 
\f {\eta}2 \right) S_{-1}(\l - \mu_j +\eta) + 
\non
&+&  \a _3^{(N)}(\l; \{z_a\}_1^N) \prod_{j=1}^M 
S_{-1} \left( \l - \mu_j + \f {3\eta}2 \right) \,, 
\eaa
where $\a_j^{(N)}(\l;\{z_a\}_1^N)=\prod_{b=1}^N 
\a_j(\l - z_b)\,; j = 1, 2, 3\,,$  
\baa
\a_1(\l) &=& \left( \l+ \eta \right) 
\left( \l+ 3\eta/2\right) \,,   
\quad \a_2(\l) = \l  \left( \l + 3\eta/2 \right) \,,  
\non
\a_3(\l) &=& \l \left( \l + \eta/2 \right) \,,  
\quad S_n(\mu) = \f {\mu - n\eta/2 }{\mu + n\eta/2 } \;. 
\eaa 
Although according to \Ref{lq} the eigenvalue has formally two 
sets of poles at $\l = \mu_j - \eta/2$ and 
$\l = \mu_j - \eta$, the corresponding residues are zero due 
to the Bethe equations \c{K85} 
\be \la{BEq} 
\prod_{a=1}^N
\left(
\f{\mu_j -z_a +  \eta/2}{\mu_j -z_a - \eta/2}\right) = 
\prod_{k=1}^M S_1(\mu_j - \mu_k) S_{-2}(\mu_j - \mu_k) \;.
\ee

If we take different spins $l_a$ at different sites of the 
lattice and the following space of states 
\ben
{\cal H} = \underset {a=1}{\overset{N}{\otimes}} V^{(l_a)}_a 
\;, 
\een
then the factors on the left hand side of \Ref{BEq} will 
be spin dependent too
\ben
\f{\mu_j -z_a +  \eta l_a/2}{\mu_j -z_a - \eta l_a/2}\;. 
\een

The \(osp(1|2)\) invariant $R$-matrix \Ref{R} has more 
complicated structure than the $sl(2)$ invariant 
$R$-matrix of C. N. Yang $R= \l I + \eta {\cal P}$. 
As a consequence the commutation relations of 
the entries $T_{ij}(\l)$  of the $T$-matrix \Ref{Tq} 
are more complicated and construction of the eigenstates 
of the transfer matrix $t(\l)$ by the algebraic Bethe Ansatz 
can be done only using a complicated recurrence relation 
expressed in terms of $T_{ij}(\mu_k)$ \c{T} 
(see also \c{M} for the case of $osp(1|2)$). It will be shown 
below that due to a simplification of this relation in the 
quasi-classical limit $\eta \to 0$ one can solve it and 
find the creation operators for the $osp(1|2)$ Gaudin model 
explicitly.


\section{$OSp(1|2)$ Gaudin model and creation operators}  

The classical $r$-matrix of the 
orthosymplectic Lie superalgebra $osp(1|2)$ can be expressed in a 
pure algebraic form using Casimir element in the tensor product 
$osp(1|2) \otimes osp(1|2)$ \c{K85} 
\be \la{cra}
{\hat r} \left( \l \right) = \frac 1{\l} 
c_2^{\otimes}  \,, 
\ee
where $c_2^{\otimes} = h \otimes h + 2 
\left( X^+ \otimes X^- + X^- \otimes X^+ \right) 
+ \left(v^+ \otimes v^- - v^- \otimes v^+ \right)$.
The matrix form of the Casimir element $\hat r$ in the fundamental 
representation $\pi$ of $osp(1|2)$ follows from \Ref{cra} by 
substituting appropriate $3 \times 3$ matrices instead of the 
$osp(1|2)$ generators \Ref{osp} and taking into account $Z_2$-graded 
tensor product of even and odd matrices \c{K85}.
Alternatively, the same matrix form of $\hat r$ can be obtained 
as a term linear in $\eta$ in the quasi-classical expansion of 
\Ref{R}
\ben 
r (\l) = \f{r_0}{\l} = \f{1}{\l}  \left( {\cal P} - K \right) \,,  
\een
where $\cal P$ is a graded permutation matrix and $K$ is a rank 
one projector. 

A quasi-classical limit  $\eta \to 0$ of the FRT-relations 
\Ref{RTT} results in a matrix form of the loop 
superalgebra relation 
\(( T(\l; \eta) = I + \eta L(\l) + {\cal O}(\eta^2))\)
\be \la{rL}  
\left[ \underset 1 {L}(\l), \, \underset 2 {L} (\mu) \right] = - 
\left[ r_{12}(\l - \mu) \, , \, \underset 1 {L}(\l) + 
\underset 2 {L}(\mu) \right] \,. 
\ee
Both sides of this relation have the usual commutators of even 
$9 \times 9$ matrices $\underset 1 {L}(\l)  = L(\l) \otimes I_3$,
$\underset 2 {L}(\mu)  = I_3 \otimes L(\mu)$ and $r_{12}(\l - \mu)$, 
where $I_3$ is $3 \times 3$ unit matrix and $L(\l)$ 
has loop superalgebra valued entries: 
\be \la{L}  
L(\l) = \left( \ba{ccc}
h(\l) & - v ^-(\l) & 2 X ^{-}(\l) \\
v ^+(\l) & 0 & v ^-(\l) \\
2 X ^+(\l) & v ^+(\l) & - h(\l) 
\ea \right) 
\ee 
The relation \Ref{rL} is a compact matrix form of the following 
commutation relations between the generators $h(\l), \, v^{\pm}(\mu), 
\, X^{\pm}(\nu)$ of the loop superalgebra under consideration
\baa \la{la}
\left[h(\l) \; , \; X^{\pm}(\mu) \right] = 
\mp \, 2 \frac {X^{\pm}(\l) - X^{\pm}(\mu)}{\l - \mu} \; & & \quad
\left[h(\l) \; , \; v^{\pm}(\mu) \right] = \phantom{2}
\mp \; \frac {v^{\pm}(\l) - v^{\pm}(\mu)}{\l - \mu} \,, \non 
\left[X ^+(\l) \; , \; X ^-(\mu) \right] = - \phantom{2}
\frac {h(\l) - h(\mu)}{\l - \mu}   \; & & \quad 
\left[X^{\pm}(\l) \; , \; v^{\mp}(\mu) \right] = - \phantom{2}
\frac {v^{\pm}(\l) - v^{\pm}(\mu)}{\l - \mu} \,, \non 
\left[v^+(\l)\; , \; v^-(\mu) \right]_+ =  \phantom {-} \phantom{2}
\frac {h(\l) - h(\mu)}{\l - \mu}  \;&&\quad 
\left[v^{\pm}(\l)\; , \; v^{\pm}(\mu) \right]_+ = 
\mp \, 2 \frac{X^{\pm}(\l) - X^{\pm}(\mu)}{\l - \mu}\,,\non 
\eaa
together with \( \left[X^{\pm}(\l) \; , \; v^{\pm}(\mu) \right] 
= 0 \).

These commutation relations \Ref{la} define the positive part 
${\cal L}_+(osp(1|2))$ of the loop superalgebra. The usual generators 
$Y_n$ of a loop algebra parameterized by non-negative integer, 
are obtained from the expansion 
$Y(\l)= \sum_{n \ge 0} Y_n \; \l^{-(n+1)}$. 
In particular, taking all $Y_n = 0$ for $n > 0$ one gets an 
$L$-operator $L(\l) = L_0 \; \l^{-1}$, where $L_0$ is $osp(1|2)$-valued 
matrix. This $L_0$ similar to $r_0$ satisfies  cubic characteristic 
equation with the $osp(1|2)$ Casimir element \Ref{Cas} as coefficient
\baa \la{L0-ev}
L ^3_0 + 2 L ^2_0 - ( c _2 - 1 ) \, L_0 - c _2 I = 0 \;.
\eaa

A Gaudin realization of the loop algebra \Ref{la} can be defined
through the generators $Y = (h, v^{\pm}, X^{\pm})$ 
\baa \la{Gr}  
Y(\l) = \sum_{a=1}^N \frac{Y_a}{\l - z_a} \;, \quad 
Y_a \in {\endo} \, (V_a)\,, 
\eaa
where $Y_a$ are \(osp (1|2) \) generators in an irreducible 
representation $V^{(l _a)}_a$ of the lowest spin $-l _a$ 
associated with each site $a$ \c{GB,S87}. Then the $L$-operator 
\Ref{L} has the form  
\baa \la{LG}  
L(\l; \{z_a\}_1^N) &=& \sum_{a=1}^N \frac{L_a}{\l - z_a}\,,  
\eaa
here $\{z_a\}_1^N$ are parameters of the model (cf \Ref{Tq}). 
It follows from the relation \Ref{LG} that the first term in the 
asymptotic expansion near $\l = \infty$ defines generators of the 
global superalgebra $osp(1|2) \subset {\cal L} _+ (osp(1|2))$ 
\baa \la{Lgl}  
L_{gl} = \lim _{\l \to \infty} \l \, L(\l) &=& \sum_{a=1}^N L_a \,,  
\eaa
with $h_{gl}$, $v ^{\pm}_{gl}$ and $X ^{\pm}_{gl}$ as generators and
entries of $L_{gl}$ (cf \Ref{L}). Moreover, from the equation 
\Ref{rL} we get
\baa \la{rLgl}  
\left[ \underset 1 {L_{gl}}, \, \underset 2 {L} (\mu) \right] = - 
\left[ r_0 \, , \, \underset 2 {L}(\mu) \right] \,, 
\eaa
here \( \underset 1 {L_{gl}} = L_{gl} \otimes I_3\), 
\eg $[h_{gl} \, , \, v ^+ (\mu) ] =  v ^+ (\mu)$.

Let us consider the loop superalgebra ${\cal L}_+(osp(1|2))$ 
as the dynamical symmetry algebra, {\ie} as the algebra of 
observables. In order to define a dynamical system besides the 
algebra of observables we need to specify a hamiltonian. It is 
a well-known fact that due to the $r$-matrix relation \Ref{rL}, 
the so-called  Sklyanin linear brackets, the elements 
\baa\la{tG} 
t(\l) &=& \frac 12 \; {\str} \; L^2(\l) = h^2(\l) + 
2 [X^+(\l)\,,\,X^-(\l)]_+ + [v^+(\l)\,,\,v^-(\l)]_- 
\non 
&=&h^2(\l) + h'(\l) + 4 X^+(\l)X^-(\l) + 2 v^+(\l)v^-(\l) 
\eaa 
commute for different values of the spectral parameter 
$t(\lambda)t(\mu) = t(\mu)t(\lambda)\,.$ Thus, $t(\lambda)$ 
can be considered as a generating function of integrals of motion. 

It is straightforward to calculate the commutation relations
between the operator $t(\l)$ and the generators of the loop algebra 
$X^+(\mu )$ and $v^+(\mu)$
\be \la{tX}   
\left [ t(\l), X^+(\mu) \right] = 4 
\f {X^+(\mu)h(\l) - X^+(\l)h(\mu)}{\l - \mu} 
- \phantom{2}\f {v^+(\l)v^+(\mu) - v^+(\mu)v^+(\l)}{\l - \mu} \;,
\ee
\be \la{tv}   
\left[ t(\l), v^+(\mu) \right] = 2 
\f {v^+(\mu)h(\l) - v^+(\l)h(\mu)}{\l - \mu} 
- 4 \f {X^+(\mu)v^-(\l) - X^+(\l)v^-(\mu)}{\l - \mu} \;.
\ee

A direct consequence of the equation \Ref{rLgl} is an invariance 
of the generating function of integrals of motion $t(\l)$ under 
the action of the global $osp(1|2)$ 
$\left[ t(\l), L_{gl} \right] = 0$.

We can consider the representation space
\({\cal H}_{ph}\) of the dynamical algebra to be a lowest spin 
\(\rho (\l)\) representation of the loop superalgebra with the 
lowest spin vector \(\Omega_-\) 
\be 
\la{vac} 
h (\l) \Om_- = \rho (\l) \Om_- \,, \quad v^-(\l) \Om_- = 0 
\,. 
\ee 
In particular, a representation of the Gaudin realization 
\Ref{LG} can be obtained by considering irreducible 
representations $V_a^{(l_a)}$ of the 
Lie superalgebra \(osp(1|2)\) defined by a spin 
$-l_a$ and a lowest spin vector $\o_a$ such that 
$v_a^-\o_a = 0$ and $h_a \o_a = -l_a \o_a$. Thus,
\be \la{Gre}
\Om _- = \mathop{\otimes} \limits_{a=1}\limits^N \o _a \;, \quad 
\hbox {and} \quad 
\rho (\l) = \sum_{a=1}^N \frac{-l_a}{\l - z_a} \;.
\ee

It is a well-known fact in the theory of GM \c{GB,S87}, 
that the Gaudin hamiltonian 
\be \la{sGh}
H^{(a)} = \sum_{b \neq a} \frac {c_2^{\otimes}(a,b)}{z_a - z_b} \;,
\ee
here $c_2^{\otimes}(a,b) = h _a h _b + 2 \left( X ^+_a X ^-_b 
+ X ^-_a X ^+_b \right) + \left( v  ^+_a v ^-_b - v ^-_a v ^+_b 
\right)$, can be obtained as the residue of the operator 
$t(\l)$ at the point $\l = z_a$ using the expansion 
\baa \la{tpole}
t(\l) &=& \sum_{a =1}^N \left( \f {l_a(l_a+1)}{(\l - z_a)^2} 
+ \, 2 \, \f {H^{(a)}}{\l - z_a} \right)  \, .
\eaa

To construct the set of eigenstates of the generating function 
of integrals of motion $t(\l)$ we have to define appropriate 
creation operators. The creation operators used in the $sl(2)$ 
Gaudin model coincide with one of the $L$-matrix entry 
\c{GB,S87}. However, in the \(osp(1|2)\) case the creation 
operators are complicated functions of the two generators 
$X^+(\l)$ and $v^+(\mu)$ of the loop superalgebra.

\begin{definition}
Let $B_{M}(\mu_1, \dots , \mu_M)$ belong to the Borel subalgebra
of the \(osp (1|2) \) loop superalgebra ${\cal L}_+(osp (1|2))$
such that
\be \la{rr}  
B_{M}(\mu_1, \dots , \mu_M) = v^+(\mu_1) 
B_{M-1}(\mu_2, \dots , \mu_M) 
+ 2 X^+(\mu_1) \sum_{j=2}^{M} 
\f {(-1)^j}{\mu_1 - \mu_j} B_{M-2}^{(j)}(\mu_2, \dots , \mu_M)  
\,, \non
\ee
with $B_0 = 1,$ $B_1(\mu) = v^+(\mu) $ and $B_M = 0$ for $M < 0$. 
The notation\break\hfil 
$B_{M-2}^{(j)}(\mu_2, \dots ,\mu_M)$ means that the argument 
$\mu_j$ is omitted.
\end{definition}

As we will show below, the $B$-operators are such that the Bethe 
vectors are generated by their action on the lowest spin vector 
$\Om _-$ \Ref{vac}. To prove this result we will need some 
important properties of the $B$-operators. 
All the properties of the creation operators 
$B_M(\mu_1, \dots , \mu_M)$ listed below can be demonstrated by 
induction method. Since their proofs are lengthy and do not contain 
illuminating insights we will omit them.

\begin{lemma}
The creation operators $B_{M}(\mu_1, \dots , \mu_M)$ are 
antisymmetric functions of their arguments
\baa \la{asym} 
B_{M}(\mu_1, \dots , \mu_{k},  \mu_{k+1}, \dots , \mu_M) &=& 
- \; B_{M} (\mu_1, \dots , \mu_{k+1}, \mu_{k}, \dots , \mu_M) 
\,, \non
\eaa
here $1\leq k < M$ and $M \ge 2$. 
\end{lemma}

Subsequently we  calculate the commutation relations between 
the generators of the loop superalgebra ${\cal L}_+(osp(1|2))$ 
and the $B$-operators. In order to simplify the formulas we 
will omit the arguments and denote the creation operator 
$B _M (\mu_1, \dots , \mu_M)$ by $B _M$.

\begin{lemma}
The commutation relations between the creation operator $B_M$
and the generators $v^+(\l)$,  $h(\l)$, $v ^-(\l)$ of the loop 
superalgebra are given by 
\be \la{vB} 
v^+(\l) B_{M} = (-1)^M B_{M} v^+(\l) + 2 \sum_{j=1}^M (-1)^j
\f {X^+(\l) -X^+(\mu _j)}{\l - \mu _j} \; B_{M-1}^{(j)} \,,
\ee
\baa \la{hB} 
h(\l) B_{M} &=&  B_{M} \left( h(\l) + \sum_{i=1}^M 
\f 1{\l - \mu_i} \right)
+ \sum _{i=1}^M \f{(-1)^i}{\l - \mu _i} \times \non
&&\times \left( v ^+(\l) B_{M-1}^{(i)}
+ 2 X ^+ (\l) \sum_{j \neq i}^M 
\f{(-1)^{j+\Th (i-j)}}{\mu _i -\mu _j} B_{M-2}^{(i,j)} 
\right) \;,
\eaa
\baa \la{v-B}
v ^-(\l)  B_{M} &=& (-1) ^M B_{M} v ^-(\l) + \sum _{j=1}^M 
\left( \f{h (\l) - h (\mu _j)}{\l - \mu _j} + \sum_{k \neq j}^M
\f{1}{(\l - \mu _k)(\mu _k - \mu _j)} \right ) \times \non 
&\times& (-1)^{j-1} B_{M-1}^{(j)} + v ^+(\l) \sum_{i < j}^M
(-1) ^{i-j-1} \f {B^{(i,j)}_{M-2}}{\mu _i - \mu _j}  
\left( \f 1{\l - \mu _i} + \f 1{\l - \mu _j} \right) \;. \non
\eaa
here the upper index of $B_{M-1}^{(j)}$ means that the argument 
$\mu_j$ is omitted, the upper index of $B_{M-2}^{(i,j)}$ means 
that the parameters $\mu_i, \mu_j$ are omitted and $\Th (j)$ 
is Heaviside function.
\end{lemma}
Already at this point we can make some useful observations.
\begin{remark}
The commutation relations between the generators of the global 
\(osp (1|2) \) and the $B$-operators follow from the previous 
property. To see this we multiply \Ref{vB}, \Ref{hB} and 
\Ref{v-B} by $\l$ and then take the limit $\l \to \infty$. 
In this way we obtain 
\baa \la{vglB} 
v^+_{gl} B_{M} &=& (-1)^M B_{M} v^+_{gl} - 2 \sum_{j=1}^M (-1)^j
X^+(\mu _j) B_{M-1}^{(j)} \;, \\
\la{hglB}
h _{gl}  B_{M} &=& B_{M} \left( h _{gl}  + M \right) \;, \\
\la{v-glB} 
v^- _{gl} B_{M} &=& (-1)^M  B_{M} \, v^-_{gl} + \sum_{j=1}^M 
(-1)^j B_{M-1}^{(j)} \left( h(\mu_j) + \sum_{k \neq j}^M 
\f 1{\mu_j - \mu_k} \right) \,. \non
\eaa 
\end{remark}
\noindent
The proof of the main theorem is based on subsequent 
property of the creation operators. 

\begin{lemma}
The generating function of integrals of motion \(t (\l)\) 
\Ref{tG} has the following commutation relation with 
the creation operator $B_M (\mu_1, \dots , \mu_M)$
\baa \la{tB}
t(\l)B_{M} &=& B_{M} t(\l) + 2 B_{M} \left( h (\l) 
\sum _{i=1}^M \f 1{\l -\mu _i} + \sum_{i < j}^M 
\f 1{(\l -\mu _i)(\l -\mu _j)} \right)
\non
&+& 2 \sum _{i=1}^M \f {(-1)^i}{\l - \mu _i} \left( v ^+(\l) 
B_{M-1}^{(i)} + 2 X ^+ (\l) \sum_{j\neq i}^M 
\f {(-1)^{j+\Th (i-j)}}{\mu _i -\mu _j} 
B_{M-2}^{(i,j)} \right) \hb _M (\mu _i) 
\non
&+& 4 \sum _{i=1}^M \frac{(-1)^i}{\l - \mu _i} 
B_{M-1}^{(i)} \left( X ^+ ( \l )  v ^- ( \mu _i ) 
- X ^+ ( \mu _i )  v ^- ( \l ) \right) \;.
\eaa
The notation we use here for the operator 
$\hb _M (\mu _i) = h(\mu_i) + \sum_{j\neq i}^M 
(\mu_i - \mu_j) ^{-1}$.
\end{lemma}

In the Gaudin realization \Ref{Gr} the creation operators 
$B_M (\mu_1, \dots , \mu_M)$ have some specific analytical 
properties.

\begin{lemma}
The $B$-operators in the Gaudin realization \Ref{Gr} satisfy 
an important differential identity
\be \la{derB}
\f {\p}{\p z_a} B _M = \sum _{j=1}^M  \f {\p}{\p \mu _j} 
\left( \f {(-1) ^j}{\mu_j - z_a} \left( v ^+_a B^{(j)}_{M-1} 
+ 2 \, X ^+_a \, \sum _{k\neq j}^M \f {(-1) ^{k+\Th (j-k)}}
{\mu_j - \mu _k}  B^{(j,k)}_{M-2} \right) \right) \;.
\ee
\end{lemma}

This identity will be fundamental step in establishing a 
connection between the Bethe vectors and the KZ equation.

The recurrence relation \Ref{rr} can be solved explicitly. 
To be able to express the solution of in a compact form 
it is useful to introduce a contraction operator $d$.

\begin{definition} Let $d$ be a contraction operator whose 
action on an ordered product 
$\overset{M}{\underset{\longrightarrow}
{\underset{j = 1}{\prod}}} v^+(\mu_j)$, $M \geq 2$, 
is given by 
\be \la{d}  
d \left(v^+(\mu_1) v^+(\mu_2) \dots  v^+(\mu_M) \right) = 
2 \sum_{j = 1}^{M-1} X^+(\mu_j) {\sum_{k=j+1}^{M}} 
\f{(-1)^{\sigma(jk)}}{\mu_j - \mu_k} 
\underset{\longrightarrow}{\prod_{m \neq j, k}^{M}} 
v^+(\mu_m) , 
\ee 
where $\s(jk) $ is the parity of the permutation 
$$ 
\s : (1, 2, \dots , j, j+1,  \dots , k, \dots , M) \to 
(1, 2, \dots , j, k, j+1, \dots , M) \;.
$$ 
\end{definition}
The $d$ operator can be applied on an ordered product
$\overset{M}{\underset{\longrightarrow}{\underset{j = 1}
{\prod}}}v^+(\mu_j)$ consecutively several times, up to 
$[M/2]$, the integer part of $M/2$. 
\begin{theorem}
Explicit solution to the recurrence relation \Ref{rr} is 
given by
\be \la{srr}  
B_{M}(\mu_1, \dots , \mu_M) = 
\underset{\longrightarrow}{\prod_{j=1}^{M}}
v^+(\mu_j) + \sum_{m = 1}^{[M/2]} \f{1}{m!} \; d^m
\underset{\longrightarrow}{\prod_{j = 1}^{M}} v^+(\mu_j) \;.  
\ee
\end{theorem}

The properties of the creation operators $B_M$ studied in 
this Section will be fundamental tools in determining 
characteristics of the $osp(1|2)$ Gaudin model. Our primary 
interest is to obtain the spectrum and the eigenvectors 
of the generating function of integrals of motion $t(\l)$ 
\Ref{tG}.

\begin{theorem}
The lowest spin vector $\Om _-$ \Ref{vac} is an eigenvector
of the generating function of integrals of motion $t(\l)$ 
\Ref{tG} with the corresponding eigenvalue $\L _0 (\l)$
\be
\la{l0-t} 
t(\l) \, \Om _- = \L _0 (\l) \, \Om _- \;, \quad \L _0 (\l)= 
\rho^2(\l) + \rho'(\l) \;.   
\ee
Furthermore, the action of the $B$-operators on the lowest 
spin vector $\Om _-$ yields the eigenvectors
\be \la{eigv}  
\Psi (\mu_1, \dots , \mu_M) = B_{M}(\mu_1, \dots , \mu_M) 
\; \Omega _- \;, 
\ee
of the $t(\l)$ operator 
\be \la{eigeq}  
t(\l) \Psi (\mu_1, \dots , \mu_M) = 
\L (\l ; \, \{\mu_j\}_{j=1}^M) \, \Psi (\mu_1, \dots , \mu_M) 
\;, 
\ee  
with the eigenvalues
\be \la{l-t} 
\L (\l ; \, \{\mu_j\}_{j=1}^M) = \L _0 (\l) + 
2 \rho (\l) \sum _{k=1}^M \f 1{\l - \mu_k} + 
2 \sum_{k < l} \f{1}{(\l - \mu_k)(\l - \mu_l)} \;, \\
\ee
provided that the Bethe equations are imposed on the 
parameters $\{\mu_j\}_{j=1}^M$
\be \la{Beq} 
\b _M (\mu_j) = \rho (\mu_j) + \sum_{k \neq j}^M 
\f {1}{\mu_j - \mu_k} = 0 \;.   
\ee 
\end{theorem}
{\it Proof.} The equation \Ref{l0-t} can be checked by a 
direct substitution of the definitions of the operator 
$t(\l)$ and the lowest spin vector $\Om _-$, 
the equations \Ref{tG} and \Ref{vac}, respectively. 

To show the second part of the theorem, we use the equation 
\Ref{eigv} to express the Bethe vectors 
$\Psi (\mu_1, \dots , \mu_M)$
\be \la{peeq}  
t(\l) \Psi (\mu_1, \dots , \mu_M) = t(\l) \; 
B_{M}(\mu_1, \dots , \mu_M) \; \Omega _- \;. 
\ee
Our next step is to use the third property of the $B$-operators, 
the equation \Ref{tB}, and the definition of
the lowest spin vector $\Om _-$ \Ref{vac} 
in order to calculate the action of the operator $t(\l)$ on the 
Bethe vectors when the Bethe equations \Ref{Beq} are imposed 
\be \la{preeq} 
t(\l)B_{M} \Om _- = B_{M} t(\l) \Om _- + 2 \left( \rho (\l) 
\sum _{i=1}^M \f 1{\l -\mu _i} + \sum_{i < j}^M 
\f 1{(\l -\mu _i)(\l -\mu _j)} \right)  B_{M} \Om _- \;.
\ee
We can express the first term on the right hand side since 
we know how the operator $t(\l)$ acts on the vector $\Om _-$, 
the equation \Ref{l0-t},
\be \la{proeeq} 
t(\l) B_{M} \Om _- = \left(\L _0 (\l) 
+ 2 \left( \rho (\l) \sum _{i=1}^M 
\f 1{\l -\mu _i} + \sum_{i < j}^M 
\f 1{(\l -\mu _i)(\l -\mu _j)} \right) \right) B_{M} \Om _- \;.
\ee
The eigenvalue equation \Ref{eigeq} as well as the expression 
for the eigenvalues \Ref{l-t} follow from the equation 
\Ref{proeeq}. 
{\qed}

\begin{corollary} \
In the Gaudin realization of the loop superalgebra given by the
equations \Ref{Gr} and \Ref{Gre} the Bethe vectors 
$\Psi (\mu_1, \dots , \mu_M)$ \Ref{eigv} are the eigenvectors of
the Gaudin hamiltonians \Ref{sGh}
\be \la{Geeq}  
H^{(a)} \Psi (\mu_1, \dots , \mu_M) = E^{(a)}_M 
\Psi (\mu_1, \dots , \mu_M) \;, 
\ee
with the eigenvalues
\be \la{GEn}
E ^{(a)}_M = \sum_{b\neq a}^{N} \f{l_a \, l_b}{z _a-z _b} 
+ \sum _{j=1}^M  \f{l_a}{\mu _j - z _a} \;,
\ee 
when the Bethe equations are imposed
\be \la{BeqG} 
\b _M (\mu_j) = \rho (\mu_j) + \sum_{k \neq j}^M 
\f {1}{\mu_j - \mu_k} = \sum_{a=1}^N 
\f {-l_a}{\mu_j - z_a} + \sum_{k \neq j}^M 
\f {1}{\mu_j - \mu_k} = 0 \,.   
\ee 
\end{corollary}
The statement of the corollary follows from residue of the
equation \Ref{eigeq} at the point $\l = z _a$. The residue 
can be determined using \Ref{tpole}, \Ref{l-t} and 
\Ref{l0-t}.

The eigenvalue \Ref{l-t} of the operator $t(\l)$ and the 
Bethe equations \Ref{Beq} can be obtained also as the 
appropriate terms in the quasi-classical limit $\eta \to 0$ 
of the expressions \Ref{lq} and \Ref{BEq}.

Comparing the eigenvalues $E^{(a)}_M$ \Ref{GEn} of the Gaudin 
hamiltonians and the Bethe equations \Ref{BeqG} with the 
corresponding quantities of the $sl(2)$ GM \c{GB,S87} 
we arrive to an interesting observation.

\begin{remark}
The spectrum of the $osp(1|2)$ Gaudin model with the spins $l_a$ 
coincides with the spectrum of the $sl(2)$ GM for the integer 
spins (cf. an analogous observation for partition functions 
of corresponding anisotropic vertex models in \c{HS}).
\end{remark}

\begin{remark}
The Bethe vectors are eigenstates of the global generator 
$h _{gl}$
\be \la{hglee}
h _{gl} \Psi (\mu_1, \dots , \mu_M) = 
\left( - \sum _{a=1}^N l_a  + M \right) 
\Psi (\mu_1, \dots , \mu_M)
\;.
\ee
Moreover, these Bethe vectors are the lowest spin vectors
of the global 
$osp(1|2)$ since they are annihilated by the 
generator $v^-_{gl}$
\be \la{vglBv} 
v^-_{gl} \Psi (\mu_1, \dots , \mu_M) = 0 \;, 
\ee
once the Bethe equations are imposed \Ref{BeqG}. These 
conclusions follow from the remark 3.2 the equations 
\Ref{hglB}, \Ref{v-glB} and the definition of the Bethe vectors
\Ref{eigv}.
\end{remark}

Hence, action of the global generator $v^+_{gl}$ on the lowest 
spin vectors $\Psi (\mu_1, \dots , \mu_M)$ generates a multiplet 
of eigenvectors of the operator $t(\l)$. One can repeat the 
arguments of \c{TF,F95} to demonstrate combinatorially
completeness of the constructed states.

As was pointed out already in \c{GB} for the $sl(2)$ case,
there are several modifications of the hamiltonians \Ref{sGh}. 
One of them is the Richardson's pairing-force hamiltonian. 
These modifications can be formulated in the framework of 
universal $L$-operator and $r$-matrix formalism (\ref{rL}) 
\c{S87}.

Due to invariance of the $r$-matrix \Ref{cra}
one can add to the $L$-operator any element of $osp(1|2)$
\baa \la{L-mod}
L (\l) \to \tilde{L} (\l) = g \, Y +  L (\l) \, ,
\eaa
preserving commutation relations \Ref{rL}. If we choose $Y = h$, 
then 
\baa \la{t-mod}
\wtt (\l) = \half \, {\str} \, \widetilde{L} ^2(\l) = 
t (\l) + 2 g \, h (\l) + g ^2
\eaa
will have the commutativity property, {\ie} 
$\wtt (\l) \wtt (\mu) = \wtt (\mu) \wtt (\l)$. 
Hence we can take $\widetilde{t} (\l)$
to be the generating function of the (modified) integrals of 
motion
\baa \la{h-mod}
\widetilde{H}^{(a)} &=& g \, h _a + \sum_{b \neq a} 
\frac {c_2^{\otimes}(a,b)}{z_a - z_b} \, .  
\eaa
Notice that the global $osp(1|2)$ symmetry is now broken 
down to global $u(1)$. In this case the eigenstates $\Psi$ 
are generated by the same B-operators, but eigenvalues and 
Bethe equations are slightly changed. Richardson like 
hamiltonian \c{GB} can be obtained as 
a coefficient of $1/\l ^2$ in the $\l \to \infty$ expansion
\baa \la{t-mod-inf}
\wtt (\l) &=& g ^2 + \f {2g}{\l} \, h _{gl} +  
\f 1{\l ^2} \, \left( 2g \, \sum _{a=1}^N z_a h _a 
+ c _2 (gl) \right) + O ( \f 1{\l ^3}) \, .
\eaa

Another modification can be obtained using an $L$-operator
with bosonic and fermionic oscillator entries
\be \la{L-osc}  
L _{osc}(\l) = \left( \begin{array}{ccc}
\l & - \g & 2 b \\
\g ^+ & 0 & \g  \\
2b ^+ & \g ^+ & - \l 
\end{array} \right) \; , 
\ee 
where $[b \, , \, b ^+] = 1$, 
$[ \g \, , \, \g ^+ ] _+ = 1$ and $\g ^2 = (\g^+ ) ^2 = 0$.
It is straightforward to see that the corresponding realization 
of the loop superalgebra will have only two nonzero commutators. 
Hence, one can consider a combination of Gaudin and oscillator 
realizations
\baa \la{L-comb}
\widetilde{L} (\l) = L _{osc}(\l) + L (\l) \;,
\eaa
preserving all the properties mentioned above of the corresponding
$B$-operators with appropriate changes in hamiltonians and Bethe
equations: $\b _M (\mu _j) \to \b _M (\mu _j) + \mu _j$.

Further modifications can be obtained considering quasi-classical
limit of the quantum spin system with non-periodic boundary 
conditions and corresponding reflection equation.

The expression of the eigenvector of a solvable model in terms of 
local variables parameterized by sites of the chain or by space 
coordinates is known as coordinate Bethe Ansatz \c{GB}. 
The coordinate representation of the Bethe vectors gives 
explicitly analytical 
dependence on the parameters $\{ \mu _i\}_1^M$ and $\{ z_a\}_1^N$
useful in a relation to the  KZ equation (Section 4).
Using the Gaudin realization \Ref{Gr} of the generators 
\ben
v^+(\mu) =  \sum_{a=1}^N \f {v^+_a}{\mu - z_a} \;, \quad 
X^+(\mu) =  \sum_{a=1}^N \f {X^+_a}{\mu - z_a} \;,  
\een
and the definition of the creation operators \Ref{srr},
one can get the coordinate representation of the 
$B$-operators: 
\be \la{coorB}  
B_{M}(\mu_1, \mu_2, ..., \mu_M) = \sum_{\pi} \left(v^+_{a_1}
\cdots v^+_{a_M} \right) _{\pi} \prod_{a=1}^N 
\vph (\{ \mu_m^{(a)}\}^{\mid \Ka _a\mid}_1 ; z_a ) \;,  
\ee
where the first sum is taken over ordered partitions $\pi$ 
of the set $(1, 2, \ldots , M)$ into subsets $\Ka _a$, 
$a= 1, 2, \ldots , N$, including empty subsets with the 
constraints
\ben
\bigcup_a \Ka _a = (1, 2, \ldots , M) \;, \quad 
\Ka _a \bigcap \Ka _b = \emptyset \quad \hbox{for} \; 
a \neq b \;.
\een
The corresponding subset of quasimomenta
\ben
\left( \mu _1^{(a)} = \mu _{j_1} , \mu _2^{(a)} = \mu _{j_2} ,
\ldots \mu _{\mid \Ka _a \mid}^{(a)} = 
\mu _{j_{\mid \Ka _a\mid }} 
; j _m \in \Ka _a \right) \;,
\een
where $\mid \Ka _a\mid$ is the cardinality of the 
subset $\Ka _a$, and $j_k<j_{k+1}$, entering into the 
coordinate wave function
\ben
\! \vph (\{ \nu _m \}_1^{\mid \Ka \mid} ; z) = \!
\sum _{\s \in {\cal S}_{\mid \Ka \mid}} (-1)^{p(\s)}
\left( (\nu_{\s(1)} - \nu_{\s(2)})  
(\nu_{\s(2)} - \nu_{\s(3)}) \cdots
(\nu_{\mid \Ka \mid} - z) \right)^{-1} .  
\een
Due to the alternative sum over permutations 
$\s \in {\cal S}_{\mid \Ka \mid}$ this functions is
antisymmetric with respect to the quasi-momenta. 
Finally the first factor in \Ref{coorB}
$\left(v^+_{a_1} \cdots v^+_{a_M} \right) _{\pi}$
means that for $j_m \in \Ka _a$ corresponding indices of 
$v^+_{a_{j_m}}$ are equal to $a$ so that  
$v^+_{a_{j_m}} = v^+_a$.
One can collect these operators into product $\prod _{a=1}^N 
\left(  v^+_a \right) ^{\mid \Ka _a \mid}$,
consequently we have an extra sign factor $(-1) ^{p(\pi)}$.

This coordinate representation is similar to the 
representations obtained in \c{BF,FFR,ReshVar} for the Gaudin 
models related to the simple Lie algebras. The $Z_2$-grading 
results in extra signs, while the complicated structure 
of the $B_M$-operators is connected with 
the fact that $(v^+_j)^2 = X_j^+ \neq 0$ while for 
$j \neq k$ $v^+_j$ and $v^+_k$ anticommute. 


\section{Solutions to the Knizhnik-Zamolodchikov equation}

Correlation functions $\psi (z _1 , \dots , z_n)$ of a two 
dimensional conformal field theory satisfy the Knizhnik-Zamolodchikov 
equation 
\c{KZ}
\be \la{KZ}
\ka \f {\p}{\p z_a} \psi (z _1 , \dots , z_n) =
\left(\sum _{b\neq a} \f {Y^{\a}_a \otimes  Y^{\a}_b}{z_a - z_b} 
\right) \psi (z _1 , \dots , z_n) \;,
\ee
where $Y^{\a}_a$ are generators of an orthonormal basis of a simple
Lie algebra in a finite dimensional irreducible representation $V_a$ 
and $\psi (z _1 , \dots , z_n)$ is a function of $N$ complex 
variables taking values in a tensor product 
$\underset {a=1}{\overset {N}{\otimes}} V_a$. The operator on the 
right hand side of \Ref{KZ} is a Gaudin hamiltonian \Ref{Gh}.

A relation between the Bethe vectors of the Gaudin model related 
to simple Lie algebras and the solutions to the
KZ equation is well known for sometime \c{BF, FFR}. Approach used 
here to obtain solutions to the KZ equation corresponding to Lie 
superalgebra $osp(1|2)$ starting from B-vectors \Ref{eigv} is 
based on \c{BF}.

A solution in question is represented as a contour integral over 
the variables $\mu_1 \dots \mu_M$
\baa \la{KZ-sol}
\psi ( z_1, \dots z_N ) &=& \oint \dots \oint 
\; \ph ( \vm | \vz ) \Psi ( \vm | \vz )
\; d\mu _1 \dots d\mu _M \; , 
\eaa
where an integrating factor $\ph ( \vm | \vz )$ is a scalar 
function
\baa \la{factor}
\ph ( \vm | \vz ) &=& \prod _{i< j}^M 
( \mu _i - \mu _j ) ^{\f 1{\ka}}
\prod _{a< b}^N ( z _a - z _b ) ^{\f {l_al_b}{\ka}}
\left( \prod _{k=1}^M \prod _{c=1}^N 
( \mu _k - z _c ) ^{\f {-l_c}{\ka}}
\right) \;, 
\eaa
and $\Psi ( \vm | \vz )$ is a Bethe vector \Ref{eigv} where 
the Bethe equations are not imposed. 
 
Differentiating the integrand of \Ref{KZ-sol} one finally gets
\be \la{final}
\ka \p _{z_a } \left( \ph  \Psi \right) = H ^{(a)} 
\left( \ph  \Psi \right) +
\ka \sum _{j=1}^M \p _{\mu_j } \left( 
\f {(-1) ^{j}}{\mu _j - z_a} \;
\ph \, {\tPsi} ^{(j,a)} \right) \;.
\ee
A closed contour integration with respect to $\mu _1 , 
\dots ,\mu _M$ will cancel the contribution from the terms 
under the sum in \Ref{final}
and therefore $\psi ( z_1, \dots z_N )$ given by \Ref{KZ-sol} 
satisfies the KZ equation. This follows from the form of the 
integrating factor \Ref{factor}
\baa
\ka \p _{z_a } \ph &=& \left(\sum _{b\neq a}^N 
\f {l_a \, l_b}{z _a-z _b} 
- \sum _{j=1}^M  \f {l_a}{z _a - \mu _j} \right) \ph 
= E ^{(a)}_M \ph \;, \\
\la{der1}
\ka \p _{\mu_j } \ph &=& \left(\sum _{a=1} ^N 
\f {-l_a}{\mu _j - z _a} 
+ \sum _{j\neq k}^M \f 1{\mu_j - \mu _k} \right) \ph 
= \b  _M (\mu _j) \ph \;,
\eaa
the differential identity \Ref{derB} and commutation 
relations \Ref{tB}.

Conjugated Bethe vectors $(B_M \Om_-) ^{\ast}$ are entering 
into solution $\tpsi( z_1, \dots z_N )$ of the dual KZ 
equation
\baa \la{dKZ}
- \ka \f {\p}{\p z_a} \tpsi (z _1 , \dots , z_N) &=& 
\tpsi (z _1 , \dots , z_N) \; H ^{(a)} \;.
\eaa
The scalar product $\left( \tpsi (z _1 , \dots , z_N) \; , 
\; \psi (z _1 , \dots , z_N) \right)$ does not depend on 
$\{z_j\} _1^N$ and its quasi-classical limit $\ka \to 0$ 
gives the norm of the Bethe vectors \c{ReshVar} due to the 
fact that the stationary points of the contour integrals for 
$\ka \to 0$ are solutions to the Bethe equations 
${\p S}/{\p \mu_j} = \b  _M (\mu _j) \ph$,
\be \la{S}
S ( \vm | \vz ) = \ka \ln \ph = \sum_{a < b}^N l_a l_b 
\ln (z_a - z_b) + \sum _{i < j}^M \ln ( \mu _i - \mu _j) 
- \sum _{a=1}^N \sum _{j=1}^M l _a \ln (z_a - \mu _j) \;. 
\ee
According to the observation in the end of Section 3 analytical 
properties of the Bethe vectors of the $osp(1|2)$ Gaudin model 
coincide with the analytical properties of the $sl(2)$ Gaudin 
model. Thus, the expression for the norm of the
Bethe vectors \Ref{eigv} obtained as the first term in the 
asymptotic expansion $\ka \to 0$ coincides also
\baa \la{norm}
&&\parallel \Psi (\mu _1 , \dots \mu _M ; \{ z_a\}_1^N ) 
\parallel ^2 =
\det \left( \f{\p ^2 S}{\p \mu_j \; \p \mu_k} \right) \;, 
\\
\f {\p ^2 S}{\p \mu_j^2} &=& \sum _{a=1}^N 
\f{l_a}{(\mu _j - z_a)^2} - 
\sum _{k \neq j}^M \f{1}{(\mu _j - \mu _k)^2} \;, 
\quad \f{\p ^2 S}{\p \mu_j \; \p \mu_k} = 
\f{1}{(\mu _j - \mu _k)^2} \;, 
\quad \hbox{for $j\neq k$} \;. \non
\eaa
Finally we notice that the modification of the Gaudin hamiltonians 
we mentioned at the end of the previous Section can be transfered 
to the corresponding modification of the KZ equations. 
Both modifications are related with extra factors in 
the integrating scalar function \Ref{factor}


\section{Conclusion} 

The Gaudin model corresponding to the simplest non-trivial Lie 
superalgebra $osp(1|2)$ was studied. A striking similarity between 
some of the most fundamental characteristics of this system and the 
$sl(2)$ GM was found. Although explicitly constructed creation 
operators $B_M$ \Ref{srr} of the Bethe vectors are complicated 
polynomials of the generators
$v^+(\l)$ and $X^+(\l)$, the coordinate form of 
the eigenfunctions differs only in signs from the corresponding 
states in the case of $sl(2)$ model. Moreover, the eigenvalues 
and the Bethe equations coincide, provided that the $sl(2)$ Gaudin 
model with integer spins is considered. 

Let us point out that using the method proposed in this paper one can 
construct explicitly creation operators of the Gaudin models 
related to trigonometric Izergin-Korepin $r$-matrix \c{T} 
and trigonometric $osp(1|2)$ $r$-matrix \c{LS}. 
Similarly to the simple Lie algebra case solutions to the 
KZ equation were constructed from the Bethe vectors 
using algebraic properties of the creation operators $B_M$ and the 
Gaudin realization of the loop superalgebra ${\cal L}_+(osp(1|2))$. 
This interplay enabled us to determine the norm of eigenfunctions
\Ref{norm}.

The difficult problem of correlation function calculation
for general Bethe vectors
\ben
{\cal C} \left( \{ \nu_j \}_1^M ; \{ \mu _i\}_1^M ; \{ \l_k\}_1^K 
\right) =
\left( \Om _- , B_M ^{\ast} (\nu _1 , \dots \nu _M ) \prod_{k=1}^K
h(\l_k) B_M (\mu _1 , \dots \mu _M ) \Om _- \right)
\een
was solved nicely for the $sl(2)$ Gaudin model in \c{S99} using the
Gauss factorization of the loop algebra group element and the
Riemann-Hilbert problem. The study of this problem 
for the $osp(1|2)$ GM is in progress and the following expression 
for the scalar product of the Bethe states is conjectured 
(cf. \c{S99})
\ben
\left( \Om _- \; , \; B_M ^{\ast} (\nu _1 , \dots \nu _M ) 
B_M (\mu _1 , \dots \mu _M ) \Om _- \right) = 
\sum _{\s \in {\cal S}_M} (-1) ^{p(\s)} \det {\cal M}^{\s} \;,
\een
where the sum is over symmetric group ${\cal S}_M$ and 
$M\times M$ matrix ${\cal M}^{\s}$ is given by
\baan
{\cal M}^{\s}_{jj} &=& 
\f{\r(\mu_j) -\r( \nu _{\s(j)})}{\mu _j - \nu _{\s(j)}}
- \sum_{k\neq j}^M 
\f{1}{(\mu _j - \mu _k)(\nu _{\s(j)} - \nu _{\s(k)})} \;, \\
{\cal M}^{\s}_{jk} &=& 
\f{1}{(\mu _j - \mu _k)(\nu _{\s(j)} - \nu _{\s(k)})} \;,
\quad \hbox{for $j,k=1, 2, \ldots M$}.
\eaan

\section{Acknowledgements}

We acknowledge useful discussions and communications with 
N. ~Yu. ~Reshetikhin, V. ~O. ~Tarasov and T. ~Takebe. This 
work was supported by the grant PRAXIS XXI/BCC/22204/99, 
INTAS grant N 99-01459 and FCT project SAPIENS-33858/99.


\end{document}